\documentclass[11pt]{elsart}
\usepackage{graphicx,color,epsfig,amsmath,amssymb}
\usepackage{subfigure}
\usepackage{cite,hyperref}
\usepackage{listings}
\graphicspath{{plots/}}
\newcounter{bla}
\newenvironment{refnummer}{%
\list{[\arabic{bla}]}%
{\usecounter{bla}%
 \setlength{\itemindent}{0pt}%
 \setlength{\topsep}{0pt}%
 \setlength{\itemsep}{0pt}%
 \setlength{\labelsep}{2pt}%
 \setlength{\listparindent}{0pt}%
 \settowidth{\labelwidth}{[9]}%
 \setlength{\leftmargin}{\labelwidth}%
 \addtolength{\leftmargin}{\labelsep}%
 \setlength{\rightmargin}{0pt}}}
 {\endlist}
\def\be{\begin{align}}
\def\ee{\end{align}}
\def\bea{\begin{align}}
\def\eea{\end{align}}
\def\nn{\nonumber}

\newcommand{\secdec}{{\textsc{SecDec}}}
\newcommand{\secdecthree}{{\textsc{SecDec}~$3$}}
\newcommand{\pysecdec}{py{\textsc{SecDec}}}
\newcommand{\fiesta}{\textsc{Fiesta}~$4.1$}
\newcommand{\python}{{\texttt{python}}}
\newcommand{\form}{{\texttt{FORM}}}

\newcommand{\eps}{\epsilon}

\begin{document}

\begin{frontmatter}
\hfill{MPP-2017-42, CERN-TH-2017-063, IPPP/17/24}\\ 

\title{pySecDec: a toolbox for the numerical evaluation of multi-scale integrals}

\author[a]{S.~Borowka},
\author[b]{G.~Heinrich},
\author[b]{S.~Jahn},
\author[b]{S.~P.~Jones},
\author[b]{M.~Kerner},
\author[c]{J.~Schlenk},
\author[b]{T.~Zirke}

\address[a]{Theoretical Physics Department, CERN, Geneva, Switzerland}
\address[b]{Max Planck Institute for Physics, F\"ohringer Ring 6, 80805 M\"unchen, Germany}
\address[c]{Institute for Particle Physics Phenomenology, University of Durham, Durham DH1 3LE, UK}

\begin{abstract}
We present \pysecdec{}, a new version  of the program \secdec{}, which performs the
factorisation of dimensionally regulated poles in parametric integrals, 
and the subsequent numerical evaluation of the finite coefficients.
The algebraic part of the program is now written in the form of \python{}
modules, which allow a very flexible usage. 
The optimization of the C++ code, generated using \form{}, 
is improved, leading to a faster numerical convergence. 
The new version also creates a library of the integrand functions, 
such that it can be linked to user-specific codes for the evaluation
of matrix elements in a way similar to analytic integral libraries.
\end{abstract}


\begin{keyword}
Perturbation theory, Feynman diagrams, multi-loop, numerical integration
\end{keyword}

\end{frontmatter}

\newpage

{\bf PROGRAM SUMMARY}
   
\begin{small}
\noindent
{\em Manuscript Title: } pySecDec: a toolbox for the numerical evaluation of multi-scale integrals                                     \\
{\em Authors: } S.~Borowka, G.~Heinrich, S.~Jahn, S.~P.~Jones, M.~Kerner, J.~Schlenk,T~.Zirke                                                \\
{\em Program Title: } pySecDec                                         \\
{\em Journal Reference:}                                      \\
{\em Catalogue identifier:}                                   \\
{\em Licensing provisions: GNU Public License v3}                                   \\
{\em Programming language:} python, FORM, C++     \\
{\em Computer:} from a single PC/Laptop to a cluster, depending on the problem\\
{\em Operating system: } Unix, Linux                                      \\
{\em RAM:} depending on the complexity of the problem                                              \\
{\em Keywords:}  Perturbation theory, Feynman diagrams, multi-loop, numerical integration
 \\
{\em Classification:}                                         
  4.4 Feynman diagrams, 
  5 Computer Algebra, 
  11.1 General, High Energy Physics and Computing.\\
 {\em External routines/libraries:} 
catch [1],
gsl [2],
numpy [3],
sympy [4],
Nauty [5],
Cuba [6],  
FORM [7], 
Normaliz [8]. 
The program can also be used in a mode which does not require Normaliz.   \\
{\em Journal reference of previous version:} Comput. Phys. Commun. 196 (2015) 470-491. \\
{\em Nature of the problem:}\\
  Extraction of ultraviolet and infrared singularities from parametric integrals 
  appearing in higher order perturbative calculations in quantum field
  theory. 
  Numerical integration in the presence of integrable singularities 
  (e.g. kinematic thresholds). \\
{\em Solution method:}\\
 Algebraic extraction of singularities within dimensional regularization using iterated sector decomposition. 
 This leads to a Laurent series in the dimensional regularization
 parameter $\epsilon$
 (and optionally other regulators), 
 where the coefficients are finite integrals over the unit-hypercube. 
 Those integrals are evaluated numerically by Monte Carlo integration.
 The integrable singularities are handled by 
 choosing a suitable integration contour in the complex plane, in an
 automated way.
 The parameter integrals forming the coefficients of the Laurent
 series in the regulator(s) are provided in the form of libraries
 which can be linked to the calculation of (multi-) loop amplitudes.
   \\
{\em Restrictions:} Depending on the complexity of the problem, limited by 
memory and CPU time.\\
{\em Running time:}
Between a few seconds and several days, depending on the complexity of the problem.\\
{\em References:}
\begin{refnummer}
\item https://github.com/philsquared/Catch/.
\item http://www.gnu.org/software/gsl/.
\item http://www.numpy.org/.
\item http://www.sympy.org/.
\item http://pallini.di.uniroma1.it/.
\item T.~Hahn, 
``CUBA: A Library for multidimensional numerical integration,''
  Comput.\ Phys.\ Commun.\  {\bf 168} (2005) 78 
  [hep-ph/0404043], 
http://www.feynarts.de/cuba/.
\item J.~Kuipers, T.~Ueda and J.~A.~M.~Vermaseren,
  ``Code Optimization in FORM,''
  Comput.\ Phys.\ Commun.\  {\bf 189} (2015) 1
  [arXiv:1310.7007], 
http://www.nikhef.nl/~form/.
\item W.~Bruns, B.~Ichim, B. and T.~R{\"o}mer, C.~S{\"o}ger, 
``Normaliz. Algorithms for rational cones and affine monoids.''
 http://www.math.uos.de/normaliz/.
\end{refnummer}
\end{small}


\section{Introduction}
\label{sec:intro}
The current experiments at the Large Hadron Collider 
as well as future collider experiments 
will explore TeV energy scales, 
posing new challenges for both the experiments and the theoretical predictions.
Radiative corrections play an important role in this situation, making
it necessary to develop calculational methods and tools which can
facilitate the task of calculating higher orders in perturbation
theory; 
both virtual (loop) corrections and real corrections, the latter of which involve extra radiation
leading to infrared singularities when unresolved.

The analytic calculation of loop integrals beyond one loop has seen an enormous progress in the last few years, 
to a large extent due to new insights~\cite{Henn:2013pwa} into the method of differential
equations~\cite{Kotikov:1991pm,Gehrmann:1999as}.
However, as the number of mass scales increases, the analytic evaluation of two-loop integrals and beyond 
becomes a very challenging and tedious task. 
At high energies, however, where massive loops are more likely to be resolved, 
and where electro-weak corrections become important, multi-scale problems are ubiquitous. 

In such cases, numerical approaches may be the method of choice. A method which is particularly useful 
in the presence of dimensionally regulated singularities is sector decomposition~\cite{Hepp:1966eg,Roth:1996pd,Binoth:2000ps,Heinrich:2008si}, 
as it provides an algorithm to factorise such singularities in an automated way.
The coefficients of the resulting Laurent series in the regularization parameter 
are in general complicated, but can be integrated numerically.
This has been implemented in the program \secdec{}~\cite{Carter:2010hi,Borowka:2012yc,Borowka:2015mxa}, 
where from version 2.0~\cite{Borowka:2012yc} the restriction to Euclidean kinematics was lifted
by combining sector decomposition with a method to deform 
the multi-dimensional integration contour into the complex plane~\cite{Soper:1999xk,Beerli:2008zz}.
Other implementations of the sector decomposition algorithm can be found in
Refs.~\cite{Bogner:2007cr,Gluza:2010rn,Ueda:2009xx,Kaneko:2010kj,Smirnov:2008py,Smirnov:2009pb,Smirnov:2013eza,Smirnov:2015mct}. 


In this paper, we present a completely new version of the \secdec{}
program, called \pysecdec{}. 
The algebraic part is now coded in \python{} and \form{}~\cite{Vermaseren:2000nd,Kuipers:2013pba} rather than Mathematica. 
As a consequence, the new program is entirely based on open source software 
and allows maximal flexibility due to its modular structure. The \python{} code writes \form{} files to produce optimized {\it C++} functions 
which can be numerically integrated with {\sc Cuba}~\cite{Hahn:2004fe,Hahn:2014fua}.
The {\it C++} functions are by default combined into a library and thus can be linked to the calculation of, for example, a full amplitude. 
Therefore \pysecdec{} can be used in a similar manner as analytic one-loop integral libraries.
This opens up new possibilities for the calculation of multi-loop amplitudes where analytic results 
for most of the master integrals are not known. 

The outline of the paper is as follows. In Section \ref{sec:program} 
we describe the structure of the program and its new features.
In Section \ref{sec:usage} we explain the installation and usage of the program.
Section \ref{sec:examples} describes a number of examples,
before we conclude in Section \ref{sec:conclusion}. 
An appendix contains an overview of possible parameter settings, where their default values are also listed.

\section{Description  of \pysecdec{}}
\label{sec:program}
\begin{figure}[htb]
\begin{center}
\includegraphics[width=12cm]{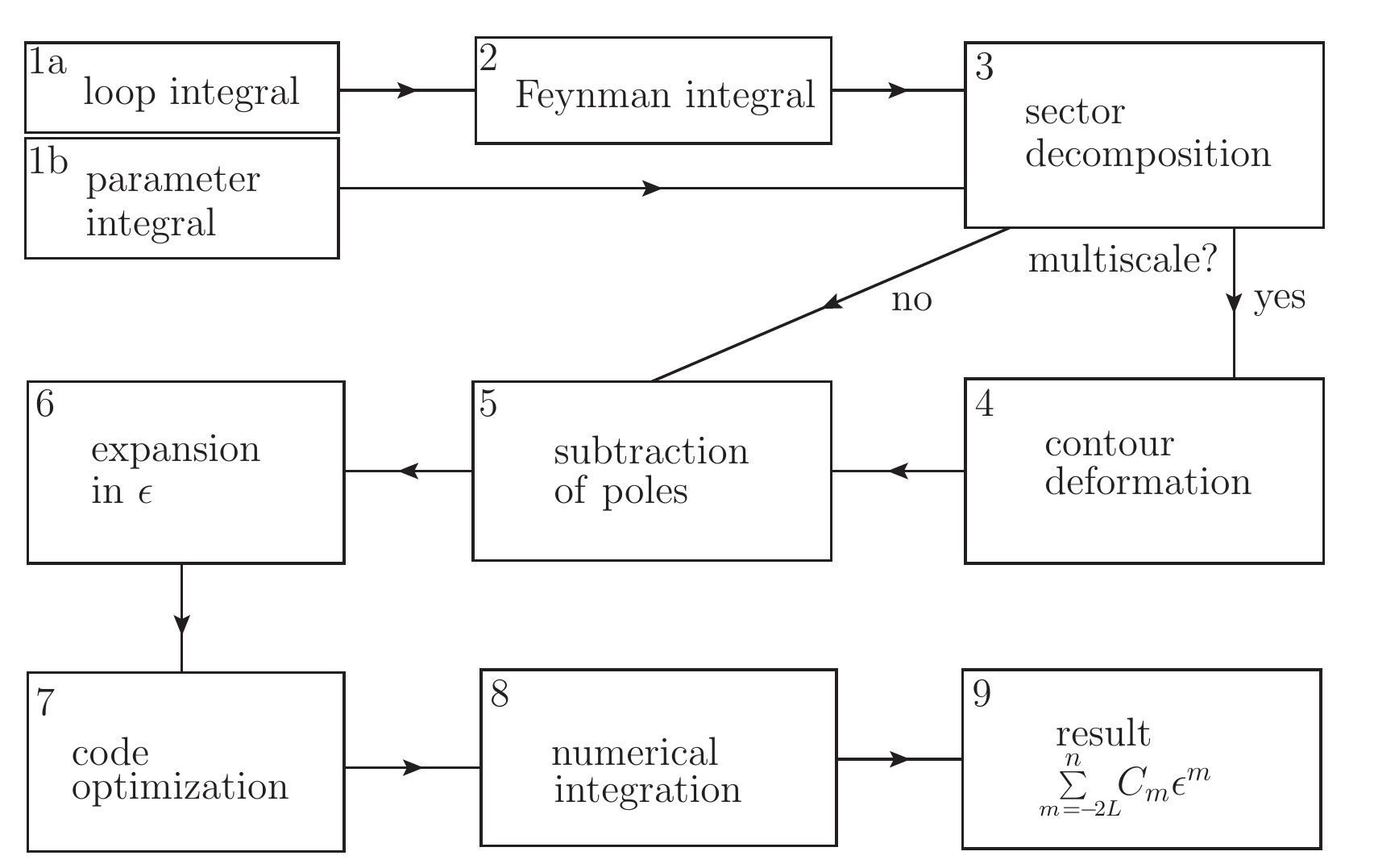}
\caption{Flowchart showing the main building blocks of
  \pysecdec{}. Steps 1-6 are done in \python{}. \form{}
  is used in step 7 to produce optimized {\it C++} code. }
\label{fig:flowchart}
\end{center}
\end{figure}

The program consists of two main parts, an algebraic part, 
based on \python{} and \form{}~\cite{Vermaseren:2000nd,Kuipers:2013pba},
and a numerical part, based on {\it C++} code. 
A flowchart is shown in Fig.~\ref{fig:flowchart}.
The isolation of endpoint singularities and the subsequent numerical integration 
can act on general polynomial functions, as indicated in the box (1b). 
Loop integrals (after Feynman parametrisation) 
can be considered as special cases of these more general polynomial integrands. 
In the \python{} code, this is reflected by the following structure: 
The \python{} function {\tt make\_package} accepts a list of polynomials raised to their individual powers as input -- corresponding to the box (1b). 
In contrast, {\tt loop\_package} takes a loop integral (e.g. from its graph or propagator representation), which corresponds to to box (1a). After Feynman parametrizing
the loop integral, {\tt loop\_package} calls  {\tt make\_package} for further processing.
The steps (1) to (6) are performed in \python{} and \form{}, where \form{}, after step (6), produces 
optimized {\it C++} code. The compiled integrand functions are by default combined into a library. For the numerical integration, 
we provide a simple interface to integrators from the {\sc Cuba}~\cite{Hahn:2004fe} library. The user also has direct access to the integrand functions 
for example to pass them to an external integrator.

The \pysecdec{} distribution comes with a very detailed documentation both in 
{\tt html} and in {\tt pdf} format.
The manual is also available in both formats on the webpage {\tt http://secdec.hepforge.org}.

\subsection{Algebraic part}

The algebraic part consists of several modules 
that provide functions and classes for the purpose of the generation of the integrand, performing the sector
decomposition, contour deformation, subtraction and expansion in the regularization parameter(s). 
The algebra modules contained in \pysecdec{} use both {\tt sympy} and {\tt numpy}, 
but also contain the implementation of a computer algebra system tailored to 
the sector decomposition purposes, in order to be competitive in speed with 
the previous implementation in Mathematica.
For example, since sector decomposition is an algorithm that acts on polynomials, a key class 
contained in {\tt pySecDec.algebra} is the class {\tt Polynomial}. 
 
Acting on general polynomials, \pysecdec{} is not limited to loop
integrals. 
It can take as an integrand any product of polynomials, raised to some
power, provided that 
the endpoint singularities are regulated by regularisation
parameters, and that integrable singularities away from the
integration endpoints can be dealt with by a deformation of the
integration path into the complex plane.
We should point out that \pysecdec{} can perform the subtraction and expansion in several regulators, 
see the description of example \ref{subsec:examples:scet}.

For loop integrals, the program contains the  module {\tt pySecDec.loop\_integral}. 
There are two ways to define a loop integral in \pysecdec{}:
(a) from the list of propagators, and (b) from the adjacency list defining the graph, 
which is roughly the list of labels of vertices 
connected by propagators. 
Examples for both alternatives to define loop integrals are given in Sections \ref{subsec:usage} and \ref{sec:examples}.

The availability of \python{} functions which can be called individually by the user allows for a very flexible 
usage of \pysecdec{}.
The {\tt html} documentation describes all the available modules and functions in detail, and also contains a ``quick search" option.

\subsubsection{Sector decomposition strategies}

When using \texttt{loop\_package}, which facilitates the definition
and calculation of Feynman integrals, one can choose between the following sector decomposition strategies:
\begin{itemize}
\item \texttt{iterative}: Default iterative method~\cite{Binoth:2000ps,Heinrich:2008si}.
\item \texttt{geometric}: Algorithm based on algebraic geometry (\texttt{G2} in \secdecthree{}). Details can be found in Refs.~\cite{Borowka:2015mxa,Schlenk:2016cwf,Schlenk:2016a}. 
\item \texttt{geometric\_ku}: Original algorithm based on algebraic geometry introduced by Kaneko and Ueda (\texttt{G1} in \secdecthree{})~\cite{Kaneko:2009qx,Kaneko:2010kj}.
\end{itemize}
The recommended sector decomposition algorithm based on algebraic geometry is \texttt{geometric}, since it improves on the original geometric algorithm.
For general parametric integrals there are additionally the following options which can be set in \texttt{make\_package}:
\begin{itemize}
\item \texttt{iterative\_no\_primary}: Iterative method without primary sector decomposition.
\item \texttt{geometric\_no\_primary}: Geometric decomposition according to Kaneko and Ueda without primary sector decomposition.
\end{itemize}



\subsection{Producing C++ code and numerical results}

The module {\tt pySecDec.code\_writer} is the main module to create a {\it C++} library.
It contains the function {\tt pySecDec.code\_writer.make\_package} 
which can decompose, subtract and expand any polynomial expression 
and return the produced set of finite functions as a {\it C++} package, 
where \form{} has been employed to write out optimised expressions.
Simple examples of how to use {\tt make\_package} are described in 
Sections \ref{subsec:examples:hypergeo} and \ref{subsec:examples:scet}.

A more advanced example is given in Section \ref{subsec:examples:dummy}, 
which shows how the user can define any number of additional finite functions.
These functions need not be  polynomial. 
Furthermore, the user is free to define arbitrary {\it C++} code (for example, a jet clustering routine) to be called by the integrand during the numerical integration. 
Templates for such user-defined functions  will be created automatically if the field 
{\tt functions}, where the names of such functions are given, is non-empty in {\tt make\_package} .

If a loop integral should be calculated, the function {\tt loop\_package} can be used, which contains methods specific to loop integrals, for example the construction of the Symanzik polynomials ${\cal F}$ and ${\cal U}$ from the list of propagators or from the adjacency list of a graph.
Examples how to use the loop package are given in Sections \ref{subsec:examples:one-loop} to \ref{subsec:examples:elliptic}.

Both {\tt make\_package} and {\tt loop\_package} will create a directory (with the name given by the user in the field {\tt name}) which contains the main {\it C++} integration files and a Makefile to generate the {\it C++} source code and the libraries 
(static and dynamic) containing the integrand functions.

The library can be linked against a user-specific code, or it can be called via a \python{} interface, as described in Section \ref{subsec:examples:one-loop}.

\subsection{New features}

In addition to the complete re-structuring, which opens up new
possibilities of usage, there are various new features compared to
\secdecthree{}: 
\begin{itemize}
\item The functions can have any number of different regulators, not
  only the dimensional regulator $\eps$.
\item The treatment of numerators is much more flexible. Numerators can
  be defined in terms of contracted Lorentz vectors or inverse
  propagators or a combination of both.
\item The distinction between ``general functions" and ``loop integrands" is removed in the sense that all features are available for both general polynomial functions and loop integrals (except those which only make sense in the loop context). 
\item The inclusion of additional ``user-defined" functions  which do not enter the decomposition has been facilitated and extended.
\item The treatment of poles which are higher than logarithmic has
  been improved.
\item A procedure has been implemented to detect and remap singularities at $x_i=1$ which cannot be cured by contour deformation.
\item A symmetry finder~\cite{2013arXiv1301.1493M} has been added which can detect isomorphisms between sectors.
\item Diagrams can be drawn (optionally, based on {\tt neato}~\cite{graphviz}; the program will however run normally if {\tt neato} is not installed).
\item The evaluation of multiple integrals or even amplitudes is now possible, using the generated {\it C++} library, as shown in Example \ref{sec:examples:4gamma}.
\end{itemize}

%
%
\section{Installation and usage}
\label{sec:usage}
Here we describe briefly the installation and usage of the
program. For more details we refer to the manual and to the examples.

\subsection{Installation}

The program can be downloaded from {\tt http://secdec.hepforge.org}.

It relies on  \python{} and runs with versions 2.7 and 3. Further
the packages {\tt numpy} ({\tt http://www.numpy.org}) and {\tt sympy}
({\tt http://www.sympy.org}) are required. The former is a  package for
scientific computing with \python{}, the latter is a \python{} library for symbolic
mathematics.
%
%
  
To install \pysecdec{}, perform the following steps

{\tt
tar -xf pySecDec-<version>.tar.gz \\
cd pySecDec-<version> \\
make \\
<copy the highlighted output lines into your .bashrc>
}

The \texttt{make} command will automatically build further dependencies
in addition to \pysecdec{} itself. These are the
{\sc Cuba} library\cite{Hahn:2004fe,Hahn:2014fua} needed for multi-dimensional numerical
integration, \form{}\cite{Vermaseren:2000nd,Kuipers:2013pba} for the algebraic manipulation of
expressions and to produce optimized {\it C++} code,
and {\sc Nauty}\cite{2013arXiv1301.1493M} to find sector symmetries, thereby reducing the total
number of sectors to be integrated.
The lines to be copied into the \texttt{.bashrc} define environment variables which make sure that
\pysecdec{} is found by \python{} and finds its aforementioned dependencies.

With our effort of shipping external dependencies with our program, we
want to make sure the installation is as easy as possible for the
user. The \pysecdec{} 
user is strongly encouraged to cite the additional
dependencies when using the program.

\subsubsection{Geometric sector decomposition strategies}
The program {\sc Normaliz}~\cite{2012arXiv1206.1916B,Normaliz} is needed for
the geometric decomposition strategies {\tt geometric} and {\tt geometric\_ku}.
In \pysecdec{}
version 1.0, the versions 3.0.0, 3.1.0 and 3.1.1 of {\sc Normaliz} are known to work.
Precompiled executables for different systems can be downloaded from \\
{\tt https://www.normaliz.uni-osnabrueck.de}. We recommend to
export its path to the environment of the terminal such that the
{\it normaliz} executable is always found. Alternatively, the path can be passed
directly to the functions that call it, see the manual for more
information.
The strategy {\tt iterative} can be used without having {\sc Normaliz} installed.

\subsection{Usage}
\label{subsec:usage}

Due to its highly modular structure, modules of the program \pysecdec{} can be
combined in such a way that they are completely tailored to the user's
needs. The individual building blocks are described in detail in the manual.
The documentation is shipped with the tarball in {\tt pdf} ({\tt
  doc/pySecDec.pdf}) and {\tt html} ({\tt doc/html/index.html}) format.

We provide \python{} scripts for the two main application directions of the
program. One is to use \pysecdec{} in a ``standalone" mode to obtain numerical results for individual integrals. This
corresponds to a large extent to the way previous \secdec{} versions were used.
The other allows the generation of a library which can be linked to the calculation of amplitudes or
other expressions, to evaluate the integrals contained in these expressions.
The different use cases are explained in detailed examples in Section~\ref{sec:examples}.

To get started, we recommend to read the section ``getting started" in the online documentation.
The basic steps can be summarized as follows:
\begin{enumerate}
\item Write or edit a \python{} script to define the integral, the replacement rules for the kinematic invariants,
the requested order in the regulator and some other options (see e.g. the one-loop box example {\tt box1L/generate\_box1L.py}).
\item Run the script using \python{}. This will generate a
  subdirectory according to the {\tt name} specified in the script.
\item Type {\tt make -C <name>}, where {\tt <name>} is your chosen name. This will create the {\it C++} libraries.
\item Write or edit a \python{} script to perform the numerical integration using the
  \python{} interface (see e.g. {\tt box1L/integrate\_box1L.py}).
\end{enumerate}
Further usage options such as looping over multiple kinematic points
are described in the documentation as well as in section \ref{subsec:examples:one-loop}.

\vspace*{5mm}

The algebra package can be used for symbolic manipulations on integrals. This can be
of particular interest when dealing with non-standard loop integrals, or if the user would like
to interfere at intermediate stages of the algebraic part.

For example, the Symanzik polynomials ${\cal F}$ and ${\cal U}$
resulting from the list of propagators can be accessed as follows
(example one-loop bubble):

{\tt >>> from pySecDec.loop\_integral import * }\\
{\tt >>> propagators = ['k**2', '(k - p)**2']}\\
{\tt >>> loop\_momenta = ['k']}\\
{\tt >>> li = LoopIntegralFromPropagators(propagators,loop\_momenta) }

Then the functions  ${\cal U}$ and ${\cal F}$ including their powers can be called as:

{\tt
>>> li.exponentiated\_U\\
( + x0 + x1)**(2*eps - 2)\\
>>> li.exponentiated\_F\\
( + (-p**2)*x0*x1)**(-eps)
}

Numerators can be included in a much more flexible way than in \secdecthree{}, see the example in Section \ref{subsec:examples:numerator} and the manual.

An example where  ${\cal F}$ and ${\cal U}$ are calculated from the
adjacency list defining a graph looks as follows (for a one-loop triangle with two
massive propagators):

{\tt
>>> from pySecDec.loop\_integral import *\\
>>> internal\_lines = [['0',[1,2]], ['m',[2,3]], ['m',[3,1]]]\\
>>> external\_lines = [['p1',1],['p2',2],['-p12',3]]\\
>>> li = LoopIntegralFromGraph(internal\_lines, external\_lines)
}
%
%

Finally, we should point out that the conventions for additional prefactors defined by the user have been changed
between \secdecthree{} and \pysecdec{}. The prefactor will now be multiplied automatically to the result.
For example, if the user defines {\tt additional\_prefactor=}$\Gamma(3-2\eps)$, this prefactor will be expanded in $\eps$
and included in the numerical result returned by \pysecdec{}.

\section{Examples}
\label{sec:examples}
All the examples listed below can be found in subdirectories of the {\tt examples} folder.

\subsection{One-loop box}
\label{subsec:examples:one-loop}

This example is located in the folder {\tt box1L}. It calculates a 1-loop box integral with one off-shell leg ($p_1^2\not=0$) and 
one massive propagator connecting the external legs $p_1$ and $p_2$.

The user has basically two possibilities to perform the numerical integrations: \\
(a) using the \python{} interface to call the library or \\
(b) using the {\it C++} interface by inserting the numerical values for the kinematic point into {\tt  integrate\_box1L.cpp}.

The commands to run this example in case (a) above are\\
{\tt 
python generate\_box1L.py\\
make -C box1L\\
python integrate\_box1L.py
}

In case (b) above the commands are\\
{\tt 
python generate\_box1L.py\\
<edit kinematic point in box1L/integrate\_box1L.cpp>\\
make -C box1L integrate\_box1L\\
./box1L/integrate\_box1L
}

The {\tt make} command can optionally be passed the jobs ({\tt -j}) command to run multiple \form{} jobs and then multiple compilation jobs in parallel, for example \\
{\tt make -j 8 -C box1L}\\
would run 8 jobs in parallel where possible.

Other simple examples can be run in their corresponding folders by replacing the name {\tt box1L} in the above commands with the name of the example.

In more detail, running the \python{} script {\tt generate\_box1L.py}
will create a folder called {\tt box1L} (the ``name'' specified in
{\tt generate\_box1L.py}) which will contain 
the following files and subdirectories:
\begin{center}  
\texttt{
\begin{tabular}{l l l l }
box1L.hpp &  integrate\_box1L.cpp & Makefile & pylink/  \\
box1L.pdf &  src/ \qquad codegen/ & Makefile.conf & README \\
\end{tabular}
}
\end{center}
Inside the generated  {\tt box1L} folder typing `{\tt make}' will create the source files for the integrand and the 
libraries `{\tt libbox1L.a}' and  `{\tt box1L\_pylink.so}', which
can be linked to an external program calling these integrals.

Note that \pysecdec{} automatically creates a {\tt pdf} file with the diagram picture if {\tt LoopIntegralFromGraph}
is used as input format. 

In case (a) the \python{} file {\tt integrate\_box1L.py} is used to perform the numerical integration, the user may edit the kinematic point and integration parameter settings directly at \python{} level.

In case (b), the user has to insert the values for the kinematic point in\\ {\tt  box1L/integrate\_box1L.cpp} at the line \\
\qquad{\tt  const std::vector<box1L::real\_t> real\_parameters = \{\}; }\\
which can be found at the beginning of {\tt int main()}.
Complex parameters should be given as a list of the real 
and imaginary part. For box1L, the complex numbers {\tt 1+2\,i} and {\tt 2+1\,i} are written as \\
\qquad {\tt const std::vector<box1L::complex\_t> complex\_parameters=\{{ \{1.0,2.0\}, \{2.0,1.0\} }\};}\\
If no complex parameters are present, the list {\tt complex\_parameters}
should be left empty. 
The command `{\tt make -C box1L box1L/integrate\_box1L}' produces the executable {\tt integrate\_box1L} which can be run to perform the numerical integration using the {\it C++} interface.

\vspace*{2mm}

{\bf Loop over multiple kinematic points}

The file {\tt integrate\_box1L\_multiple\_points.py} shows how
to integrate a number of kinematic points sequentially.
The points are defined in the file {\tt kinematics.input}. They are read in
by the \python{} script (line by line).\\
The first entry of each line in the kinematics data file {\tt
  kinematics.input} should be a string, the ``name'' of the kinematic
point, which can serve to label the results for each point. The results are written 
to the file {\tt results\_box1L.txt}.

\subsection{Two-loop three-point function with massive propagators}

\begin{figure}[htb]
\begin{center}
\includegraphics[width=5.5cm]{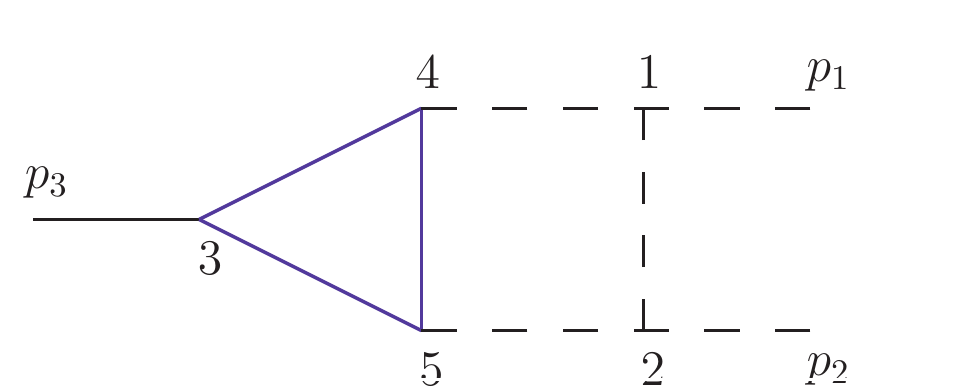}
\end{center}
\caption{A two-loop vertex graph containing a massive triangle loop.  
Solid lines are massive, dashed lines are massless. The vertices are labeled to match the 
construction of the integrand from the topology.}
\label{fig:P126}
\end{figure}

The example {\tt triangle2L} 
calculates the two-loop diagram shown in Fig.~\ref{fig:P126}.
The steps to perform to run this example (using the python interface)
are  analogous to the ones given in the previous section:\\
{\tt python generate\_triangle2L.py \&\& make -C triangle2L \&\& }\\
{\tt python integrate\_triangle2L.py}

Results for this diagram can be found e.g. in \cite{Fleischer:1997bw,Davydychev:2003mv,Bonciani:2003hc,Ferroglia:2003yj}.

The result  of \pysecdec{} is shown in Tab.~\ref{tab:triangle2L}.
We would like to point out that the default accuracy in this example
is set to $10^{-2}$ in order to keep the runtimes low.
This does not reflect the accuracy \pysecdec{} can actually reach.

\begin{table}[htb]
\caption{Result for the two-loop triangle {\tt P126} at $p_3^2=9$ and
  $m^2=1$ compared to the analytic result of Ref.~\cite{Davydychev:2003mv}. }
\begin{center}  
\begin{tabular}{|l|l|c|c|}
\hline
$\epsilon$ order & \pysecdec{} result\\
\hline
$\epsilon^{-2}$ & (-0.0379735 - i\,0.0747738) $\pm$ (0.000375449 + i\,0.000695892) \\
$\epsilon^{-1}$ & (0.2812615 + i\,0.1738216) $\pm$ (0.003117778 + i\,0.002358655) \\
$\epsilon^{0}$ & (-1.0393673 + i\,0.2414135) $\pm$ (0.011940978 + i\,0.004604699) \\
\hline
& analytic result\\
\hline
$\epsilon^{-2}$& -0.038052884394 - i\,0.0746553844162\\
$\epsilon^{-1}$& \,0.279461083591 + i\,0.1746609123993\\
$\epsilon^{0}$ &-1.033851309109 + i\,0.2421265865644\\
\hline
\end{tabular}
\end{center}
\label{tab:triangle2L}
\end{table}
 
A comparison of the timings with \secdecthree{} and \fiesta{} for the evaluation of the finite part can
be found in Table~\ref{tab:timings}.

\subsection{Two-loop four-point function with numerators}
\label{subsec:examples:numerator}

The example {\tt box2L\_numerator} shows how numerators can be treated in \pysecdec{}.
It calculates a massless planar on-shell two-loop 7-propagator box in two different ways: \\
(a) with the numerator defined as an inverse propagator ({\tt box2L\_invprop.py}), \\
(b) with the numerator defined in terms of contracted Lorentz vectors\\ ({\tt box2L\_contracted\_tensor.py}). 

This example is run with the following commands\\
\begin{tabular}{ll}
{\tt make } &{\it  (will run both \python{} scripts and compile the libraries)}\\
{\tt./integrate\_box2L} & {\it  (will calculate the integral in both
  ways and }\\
& {\it print the results 
as well as the difference between }\\
& {\it the two results, which should be numerically zero).}
\end{tabular}

The result for $s=-3$ and $t=-2$ is listed in Tab.~\ref{tab:box2l}.

The integral is given by
\begin{align}
&I_{a_1\ldots a_8}=\int\frac{d^Dk_1}{i\pi^\frac{D}{2}}\frac{d^Dk_2}{i\pi^\frac{D}{2}}
\frac{1}{[D_1]^{a_1}[D_2]^{a_2}[D_3]^{a_3}[D_4]^{a_4}[D_5]^{a_5}[D_6]^{a_6}[D_7]^{a_7}[D_8]^{a8}}\\
&D_1=k_1^2, D_2=(k_1+p_2)^2, D_3=(k_1-p1)^2,D_4=(k_1-k_2)^2,\nn\\
& D_5=(k_2+p_2)^2,D_6=(k_2-p_1)^2, D_7=(k_2+p_2+p_3)^2,D_8=(k_1+p_3)^2.\nn
\end{align}
In case (a), the integral $I_{1111111-1}$ is specified by {\tt powerlist = [1,1,1,1,1,1,1,-1]} in {\tt box2L\_invprop.py}.\\
In case (b), the same integral is specified (in {\tt box2L\_contracted\_tensor.py}) \\
by {\tt powerlist = [1,1,1,1,1,1,1,0]} and \\{\tt numerator = `k1(mu)*k1(mu) + 2*k1(mu)*p3(mu) + p3(mu)*p3(mu)'}.

\begin{table}[htb]
\caption{Result for the two-loop four-point function with numerators at the kinematic point $s=-3, t=-2$. }
\begin{center} 
\begin{tabular}{|c|c|}
\hline
$\epsilon$ order & \pysecdec{} result \\
\hline
$\epsilon^{-4}$ & -0.2916 $\pm$ 0.0022 \\
$\epsilon^{-3}$ & 0.7410 $\pm$ 0.0076 \\
$\epsilon^{-2}$ & -0.3056 $\pm$ 0.0095 \\
$\epsilon^{-1}$ & -2.2966 $\pm$ 0.0313 \\
$\epsilon^{0}$ & 1.1460 $\pm$ 0.0504 \\
\hline
\end{tabular}
\end{center}
\label{tab:box2l}
\end{table}

\subsection{Three-loop triangle integral}
\label{subsec:examples:3loop}

\begin{figure}[h]
\begin{center}
\includegraphics[width=5.3cm]{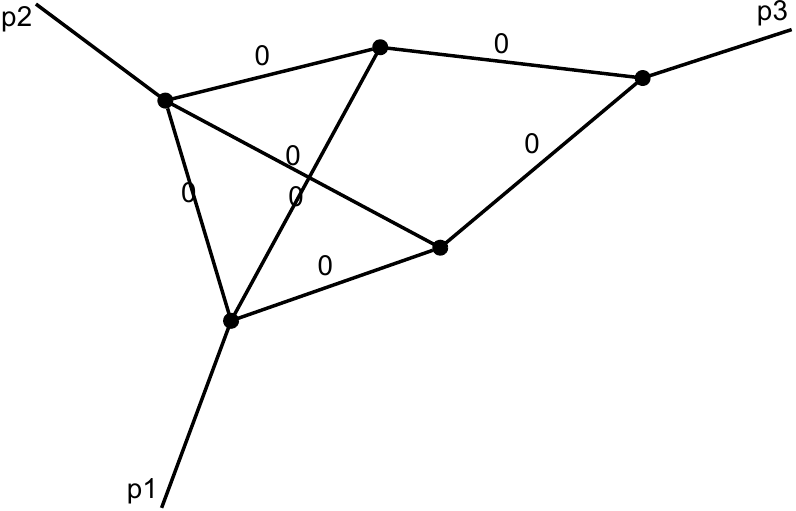}
\end{center}
\caption{Three-loop massless 7-propagator graph.
\label{fig:A75}}
\end{figure}

The example {\tt triangle3L} demonstrates how the symmetry finder can reduce the number of sectors. 
We consider the seven-propagator 3-loop 3-point integral depicted in Fig.~\ref{fig:A75}, which is the figure that is automatically created by the code.
This integral has been calculated to order $\eps$ 
in Ref.~\cite{Heinrich:2007at} and  to order $\eps^4$ in  Ref.~\cite{vonManteuffel:2015gxa}. 
Here we also calculate it to order $\eps^4$.  

This example is run as usual by the commands\\
{\tt python generate\_triangle3L.py \&\& make -C  triangle3L \&\& }\\
{\tt python integrate\_triangle3L.py}

It shows that the symmetry finder reduces the number of primary sectors to calculate from 7 to 3, 
and the total number of sectors from 212 to 122.
For comparison, \secdecthree{} produces 448 sectors using strategy X. 

\subsection{Integrals containing elliptic functions}
\label{subsec:examples:elliptic}

\begin{figure}[htb]
\begin{center}
\includegraphics[width=5.cm]{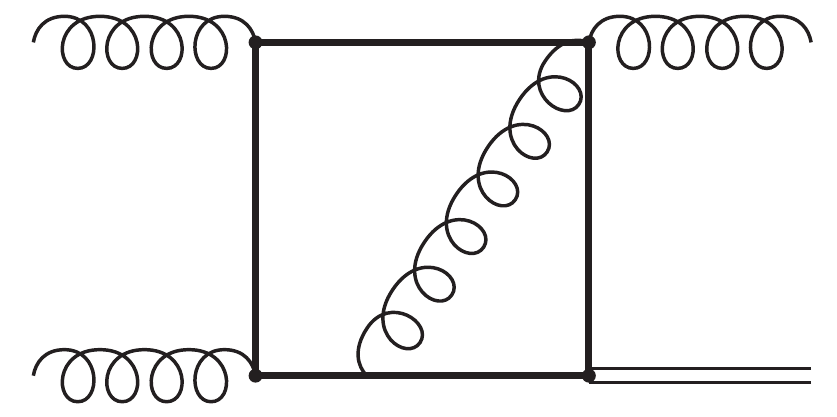}
\end{center}
\caption{Two-loop 6-propagator graph leading to elliptic functions. Curly lines denote massless particles.
The box contains massive propagators with mass $m$. One leg ($p_4$) has $p_4^2\not=0$.
\label{fig:ellipticI1}}
\end{figure}

In the examples {\tt elliptic2L\_euclidean}  and {\tt  elliptic2L\_physical} an integral is calculated which is known from 
Refs.~\cite{Bonciani:2016qxi,Primo:2016ebd}
to contain elliptic functions. 
We consider the integrals $I_{a_1\ldots a9}$ 
\begin{align}
&I_{a_1\ldots a_9}=\int\frac{d^Dk_1}{i\pi^\frac{D}{2}}\frac{d^Dk_2}{i\pi^\frac{D}{2}}
\frac{D_8^{-a8}D_9^{-a9}}{[D_1]^{a_1}[D_2]^{a_2}[D_3]^{a_3}[D_4]^{a_4}[D_5]^{a_5}[D_6]^{a_6}[D_7]^{a_7}}\\
&D_1=k_1^2-m^2, D_2=(k_1+p_1+p_2)^2-m^2, D_3=k_2^2-m^2,\nn\\
& D_4=(k_2+p_1+p_2)^2-m^2,D_5=(k_1+p_1)^2-m^2,D_6=(k_1-k_2)^2,\nn\\
& D_7=(k_2-p_3)^2-m^2,D_8=(k_2+p_1)^2,D_9=(k_1-p_3)^2.\nn
\end{align}
The topology for $I_{110111100}$ is depicted in Fig.~\ref{fig:ellipticI1}.
Here we calculate the  integral $f^A_{66}=\left(-s/m^2\right)^\frac{3}{2}\,I_{110111100}$ 
discussed in Ref.~\cite{Bonciani:2016qxi}. 

In {\tt elliptic2L\_euclidean} we calculate
the kinematic point
$s=-4/3, t= -16/5, p_4^2=-100/39, m=1$ (Euclidean point) with the settings {\tt epsrel=}$10^{-5}$, 
{\tt maxeval=}$10^7$ and obtain
\begin{align}
f^A_{66}&=0.2470743601 \pm  6.9692\times 10^{-6}  \;.
\end{align}
The analytic result\footnote{We thank Francesco Moriello and Hjalte Frellesvig for providing us the result.} is given by 
\begin{align}
f^A_{66,{\mathrm{analytic}}}&=0.247074199140732131068066 \;.\nn
\end{align}

In {\tt elliptic2L\_physical} we calculate the non-Euclidean point 
$s=90, t= -2.5, p_4^2=1.6, m^2=1$ and find with {\tt epsrel=}$10^{-4}$, 
{\tt maxeval=}$10^7$:
\begin{align}
\left(\frac{-s}{m^2}\right)^{-\frac{3}{2}}\,f^A_{66}&=-0.04428874+i\,0.01606818 \pm (2.456+i\,2.662)\times 10^{-5})\;.\nn
\end{align}

\subsection{Two-loop vertex diagram with special kinematics}

In the example {\tt triangle2L\_split} we calculate an integral entering the  
two-loop corrections to the $Zb\bar{b}$ vertex, calculated in Refs.~\cite{Fleischer:1998nb,Dubovyk:2016zok}, where it is called $N_3$.  

This example is run as usual by the commands\\
{\tt python generate\_triangle2L\_split.py \&\& make -C
  triangle2L\_split \&\&}\\
{\tt  python integrate\_triangle2L\_split.py}

\begin{figure}[htb]
\begin{center}
\includegraphics[width=0.4\textwidth]{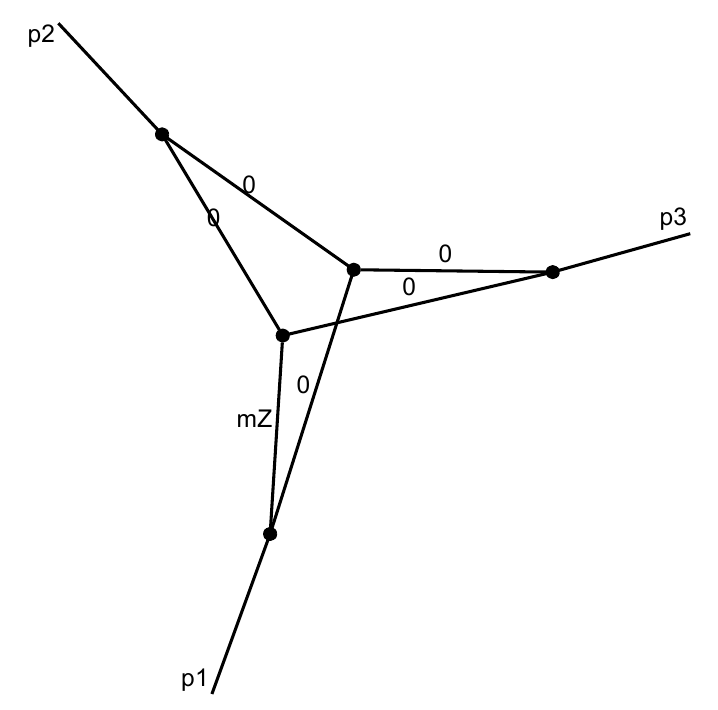}
\end{center}
\caption{The integral $N_3$ with one massive propagator ($m_Z$) and $s=p_3^2=m_Z^2$. 
\label{fig:Zbb}}
\end{figure}

The diagram produced by \pysecdec{} is shown in Fig.~\ref{fig:Zbb}.

The kinematic condition $s=M_Z^2$ leads to an integrand which is particularly difficult for the sector decomposition method 
because it does not have a Euclidean region. As a consequence, 
the integral has both endpoint singularities as well as singularities due to the fact that the second Symanzik polynomial ${\cal F}$ 
can vanish on some hyperplane in Feynman parameter space, rather than only at the origin.
The remappings done by the standard sector decomposition algorithm would turn this into singularities at $x_i=1$.
In \secdecthree, singularities at $x_i=1$ were treated by a split of the integration domain at $x_i=0.5$ and subsequent remapping 
to the unit hypercube. However, this can lead to an infinite recursion of the problem.

\pysecdec{} can detect and remap such ``hyperplane singularities" into singularities at the origin by a dedicated spitting procedure, 
where a splitting at the symmetric point $x_i=0.5$ is avoided.

The results obtained for this example are listed in Table~\ref{tab:N3}.

\begin{table}[htb]
\caption{Numerical result from \pysecdec{} for the integral $N_3$.}
\begin{center} 
\begin{tabular}{|c|l|}
\hline
$\epsilon$ order &  \pysecdec{} result  \\
\hline
$\epsilon^{-2}$ & (1.23370112 + i\,5.76 $\times 10^{-7}$) $\pm$ (0.00003623 + i\,0.00003507) \\
$\epsilon^{-1}$ & (2.89050847 + i\,3.87659429) $\pm$ (0.00060165 + i\, 0.00070525) \\
$\epsilon^0$ & (0.77923028 + i\,4.13308243) $\pm$ (0.00815782 + i\, 0.00923315)\\
\hline
\end{tabular}
\end{center}
\label{tab:N3}
\end{table}

\subsection{Hypergeometric function $_5F_4$}
\label{subsec:examples:hypergeo}

An example of a general dimensionally regulated parameter integral, which can also 
have endpoint-singularities at $z_i=1$,  can be found in {\tt hypergeo5F4}.
We consider the hypergeometric function $_5F_4(a_1,...,a_5;b_1,...,b_4;\beta)$.

This example is run by the usual commands\\
{\tt python generate\_hypergeo5F.py \&\& make -C  hypergeo5F4 \&\& }\\
{\tt python integrate\_hypergeo5F.py}

The considered function has the integral representation
$$\prod_{i=1}^4 \left[\frac{\Gamma[b_i]}{\Gamma[a_i]\Gamma[b_i-a_i]}\int_0^1dz_i\,(1 - z_i)^{-1 -
a_i + b_i}z_i^{-1 + a_i}\right](1 - \beta z_1 z_2 z_3 z_4)^{a_5}\;.$$
The potential singularities at $z_i=1$ are automatically detected by
the program and remapped to the origin 
if the flag {\tt split=True} is set.
Results for values
$a_5=-\epsilon,a_2=-\epsilon,a_3=-3\epsilon,a_4=-5\epsilon,a_5=-7\epsilon$,
$b_1=2\epsilon,b_2=4\epsilon,b_3=6\epsilon,b_4=8\epsilon, \beta=0.5$ are shown
in Tab.~\ref{tab:hyp5f4}.

\begin{table}[h!]
\caption{Comparison of the exact result for $_5F_4$ with the evaluation of \pysecdec{}, maximally using $10^9$ integrand evaluations. }
\begin{center} 
\begin{tabular}{|c|c|c|}
\hline
$\epsilon$ order & Exact result (using HypExp~\cite{Huber:2005yg}) & \pysecdec{} result  \\
\hline
$\epsilon^0$ & 1 & $1\pm\times 10^{-15}$ \\
$\epsilon^1$ & 0.1895324 & 0.18953239 $\pm$ 0.0002 \\
$\epsilon^2$ & - 2.2990427 & -2.2990377 $\pm$ 0.0016 \\
$\epsilon^3$ & 55.469019 & 55.468712 $\pm$ 0.084 \\
$\epsilon^4$ & - 1014.3924 & -1014.3820 $\pm$ 0.89 \\ 
\hline
\end{tabular}
\end{center}
\label{tab:hyp5f4}
\end{table}

\subsection{Function with two different regulators}
\label{subsec:examples:scet}

The example {\tt two\_regulators} demonstrates the sector decomposition
and integration of a function with multiple regulators.
We consider the integral\footnote{We thank Guido Bell and Rudi Rahn
  for providing this example.}
\begin{eqnarray} 
I&=e^{-\gamma_E(2\eps+\alpha)}\int_0^1 dz_0 \int_0^1 dz_1
\,z_0^{-1-2\eps-\alpha}(1-z_0)^{-1+2\eps+\alpha}
z_1^{-\eps+\frac{\alpha}{2}}\,e^{-z_0/(1-z_0)}\\
&=\frac{2}{\alpha}\Gamma(-2\eps-\alpha)\,e^{-\gamma_E(2\eps+\alpha)}\nn\\
&=-\frac{1}{\alpha\eps}+\frac{1}{2\eps^2}-\frac{\pi^2}{6}+{\cal O}(\alpha,\eps)\;.\nn
\end{eqnarray}

This example can be run using the usual commands\\
{\tt python generate\_two\_regulators.py \&\& make -C two\_regulators
  \&\&}\\
{\tt  python integrate\_two\_regulators.py}

The regulators are specified in a list as {\tt regulators =
  [`alpha',`eps']}. The orders to be calculated in each regulator are
defined in a list where the position of each entry matches the one in
the regulator list. For example, if the integral should be calculated
up to the zeroth order in the regulator $\alpha$ and to first order in
$\eps$, the corresponding input would be {\tt requested\_orders = [0,1]}.

\subsection{User-defined additional functions}
\label{subsec:examples:dummy}

The user has several possibilities to define functions  which are not included in the decomposition procedure itself 
and which can therefore be non-polynomial or be defined by an arbitrary {\it C++} function, 
for example  a jet algorithm or the
definition of an event shape variable.
Three examples ({\tt dummyI}, {\tt dummyII} and {\tt thetafunction}) which demonstrate the use of user defined functions are contained in the subdirectory {\tt userdefined\_cpp}.

\subsubsection{Analytic functions not entering the decomposition}

The example {\tt dummyI} demonstrates how a result can be multiplied by an analytic function of the integration variables which should not be decomposed. The example can be run with the usual commands\\
{\tt python generate\_dummyI.py \&\& make -C dummyI \&\& \\
python integrate\_dummyI.py}

The functions which are to be multiplied onto the result are listed in {\tt generate\_dummyI.py}
on the line {\tt functions = [`dum1', `dum2']}.
The user can give functions any name which is not a reserved \python{} function name.\\
The dependence of these functions on a number of arguments is given on the line \\
{\tt remainder\_expression =\\
`(dum1(z0,z1,z2,z3) + 5*eps*z0)**(1+eps) *\\
dum2(z0,z1,alpha)**(2-6*eps)'}.\\
Note that the {\tt remainder\_expression} is an explicitly defined function of the integration variables but
that the functions {\tt dum1} and {\tt dum2} are left implicit in the \python{} input file.

Any functions left implicit in the \python{} input (in this example {\tt dum1} and {\tt dum2})
are to be defined later in the file {\tt <name>/src/functions.hpp}.
A template for this file will be created automatically together with the process directory.
In our example, for the user's convenience, the appropriate functions are copied to the process directory in the last line of {\tt generate\_dummyI.py}.
Note that the arguments in {\tt functions.hpp} are the ones that occur in the argument list of the function in {\tt generate\_dummyI.py}, in the same order.
The function arguments can be both integration variables and parameters. 
Derivatives of the functions are needed if higher than logarithmic poles appear in the decomposition of the integrand. 
The definition of the derivatives are named following the pattern {\tt d<function>d<argument>}, 
for example `{\tt ddum1d0}' means the first derivative of the
function with name `{\tt dum1}' with respect to its first argument.

Alternatively, if the extra functions are simple, they can be defined explicitly in the \python{} input file 
in {\tt remainder\_expression =} `{\it define explicit function here}'.
The example {\tt dummyII} demonstrates this. It can be run with the usual commands\\
{\tt python generate\_dummyII.py \&\& make -C dummyII \&\& \\
python integrate\_dummyII.py}

In this case, the definition of functions like {\tt dum1,dum2} is obsolete.
The definitions given in {\tt remainder\_expression} will be multiplied verbatim to the polynomials to decompose.

\subsubsection{Non-analytic or procedural functions not entering the decomposition}

The user can also multiply the result by {\it C++} functions which are not simple analytic functions, 
for example they may contain {\tt if} statements, {\tt for} loops, etc., as may be needed
to define measurement functions or observables. An example of this is given in {\tt generate\_thetafunction.py} 
which shows how a theta-function can be implemented in terms of a {\it C++} {\tt if} statement. 
This example can be run with the usual commands\\
{\tt python generate\_thetafunction.py \&\& make -C thetafunction \&\& \\
python integrate\_thetafunction.py}

In the \python{} input file the name of the {\it C++} function is given on the line
{\tt  functions = ['cut1']}. The line\\
{\tt remainder\_expression = `cut1(z1,delt)'}\\
instructs \pysecdec{} to multiply the function onto the result, without decomposition.
Note that the implementation of the function {\tt cut1} is not given in the \python{} input file. 
Once the process directory is created, the function {\tt cut1} should be defined in {\tt <name>/src/functions.hpp}. 
In our example, the appropriate function is copied to the process directory in the last line of
{\tt generate\_thetafunction.py} for the user's convenience.
The theta-function may be defined as follows:\\

\lstset{language=C++,
                basicstyle=\ttfamily,
                keywordstyle=\ttfamily,
                stringstyle=\ttfamily,
                commentstyle=\ttfamily
}
\begin{lstlisting}
    template<typename T0, typename T1>
    integrand_return_t cut1(T0 arg0, T1 arg1)
    {
        if (arg0 < arg1) {
            return 0.;
        } else {
            return 1.;
        }
    };
\end{lstlisting}

The first argument ({\tt arg0}) corresponds to {\tt z1}, the second one ({\tt arg1}) is the cut parameter {\tt delta}.

\subsection{Four-photon amplitude}
\label{sec:examples:4gamma}

This example, contained in {\tt 4photon1L\_amplitude}, calculates the one-loop four-photon amplitude 
${\cal M}^{++--}$. The example may be run using the commands:\\
{\tt make \&\& ./amp}

The {\tt Makefile} will produce the libraries for the two-point and four-point functions entering the amplitude and compile the file {\tt amp.cpp} which defines the amplitude. Executing `{\tt ./amp}' evaluates the amplitude numerically and prints the analytic result for comparison.

The amplitude for 4-photon scattering via a massless fermion loop can be expressed in terms of three independent helicity amplitudes, 
${\cal M}^{++++}$, ${\cal M}^{+++-}$, ${\cal M}^{++--}$, out of which the 
remaining helicity amplitudes forming the full amplitude can be reconstructed using crossing symmetry, Bose-symmetry and parity. Omitting an overall factor of $\alpha^2$, 
the analytic expressions read (see e.g.~\cite{Binoth:2002xg})

\begin{eqnarray}
{\cal M}^{++++} &=& 8 \quad , \quad 
{\cal M}^{+++-} = -8 \;,\nn \\
{\cal M}^{++--} &=& -8 \Bigl[  1 + \frac{t-u}{s} \log\left(\frac{t}{u}\right) 
 + \frac{t^2+u^2}{2 s^2} \Bigl( \log\left(\frac{t}{u}\right)^2 + \pi^2 \Bigr)\Bigr] \;\; .
\end{eqnarray}
Up to an overall phase factor,
the amplitude ${\cal M}^{++--}$ can be expressed in terms of one-loop integrals as
\begin{align}
{\cal M}^{++--} =-8 
 \left\{ 1 + \frac{t^{2}+u^{2}}{s} I_{4}^{D+2}(t,u) +  \frac{t-u}{s}\left(
    I_{2}^{D}(u)-I_{2}^{D}(t) \right) \right\}\,.
\end{align}
The purpose of our simple example is to show how \pysecdec{} can be used to calculate the master integrals occurring in the amplitude ${\cal M}^{++--}$.  

\subsection{Comparison of timings}

\begin{table}[htb]
\caption{Comparison of timings (algebraic, numerical) using \pysecdec{}, \secdecthree{} and \fiesta{}.}
\begin{center} 
\begin{tabular}{|c|c|c|c|c|}
\hline
& \pysecdec{}   time\,(s) & \secdecthree{}  time\,(s) & \fiesta{} time\,(s) \\
\hline
\texttt{triangle2L} 				& (40.5, 9.6) 		& (56.9, 28.5) 			& (211.4, 10.8) \\
\texttt{triangle3L} 		& (110.1, 0.5) 	& (131.6, 1.5) 			& (48.9, 2.5) \\
\texttt{elliptic2L\_euclidean} 						& (8.2, 0.2) 		& (4.2, 0.1) 				& (4.9, 0.04) \\
\texttt{elliptic2L\_physical}			& (21.5, 1.8) 		& (26.9, 4.5) 				& (115.3, 4.4) \\
\texttt{box2L\_invprop} 	& (345.7, 2.8) 	& (150.4, 6.3) 			& (21.5, 8.8) \\
\hline
\end{tabular}
\end{center}
\label{tab:timings}
\end{table}

We compare the timings for several of the above mentioned examples between \pysecdec{}, \secdecthree{} and \fiesta{},
where we distinguish between the time needed to perform the algebraic
and the numeric part. 
In Tab.~\ref{tab:timings},
the compilation of the generated {\it C++} functions is included in the
algebraic part, because it needs to be done only once. The timings for
the numerical part are the wall clock times for the evaluation of the {\it C++} functions.

The timings were taken on a four-core (eight hyper-thread) Intel(R) Core(TM) i7-4770 CPU @ 3.40GHz
machine.  We set the parameter \texttt{number\_of\_presamples} in
\pysecdec{}, \texttt{optlamevals} in \secdecthree{} and
\texttt{LambdaIterations} in \fiesta{}, which controls the number of
samples used to optimise the contour deformation, to the \fiesta{}
default of \texttt{1000}. The default decomposition strategy of each
tool was used, \texttt{STRATEGY\_S} for \fiesta{} and \texttt{X} for
\pysecdec{} and \secdecthree{}. The integrands were summed before
integrating in the following way:
setting \texttt{together=True} in \pysecdec{} and
\texttt{togetherflag=1} in \secdecthree{} sums all integrands
contributing to a certain pole coefficient before integrating.
\texttt{SeparateTerms=False} in \fiesta{} sums the integrands in each
sector which appears after pole resolution before integrating.
For the examples considered on our test platform these settings were found to be optimal for all three tools.
The integration is performed using the default settings of \pysecdec{} and the
same settings in \secdecthree{} and \fiesta{}. In
particular, this implies a rather low desired relative accuracy of $10^{-2}$. 

The numerical integration times in \pysecdec{} are generally reduced with respect to \secdecthree{}, which
is mostly due to a better optimization during the algebraic part, and
partly also due to a more efficient deformation of the integration contour in \pysecdec{}.
For the test cases considered we found that \fiesta{} is the fastest to perform the algebraic (decomposition) step when contour deformation is not required. We would like to stress that although we endeavoured to keep all relevant settings identical across the tools we are not experts in the use of \fiesta{} and we expect that it is possible to obtain better timings by adjusting settings away from their default values. Furthermore, which tool is fastest strongly depends on the case considered and whether one prefers faster decomposition or numerical evaluation of the resulting functions.

\vspace*{3mm}

\section{Conclusions}
\label{sec:conclusion}
We have presented a new version  of the program \secdec{},  called
\pysecdec{}, which is publicly available at {\tt http://secdec.hepforge.org}.
The program \pysecdec{} is entirely based on open source software (\python{}, \form{}, {\sc Cuba}) 
and can be used in various contexts. 
The algebraic part can isolate poles in any number of regulators from general polynomial expressions,
where Feynman integrals are a special case of.
For the numerical part, a library of {\it C++} functions is created, which allows very flexible usage, 
and in general outperforms \secdecthree{} in the numerical evaluation times.
In particular, it extends the functionality of the program from the evaluation of individual (multi-)loop integrals  
to the evaluation of larger expressions containing multiple analytically unknown integrals, as for example two-loop amplitudes. 
Such an approach already has been used successfully for the two-loop integrals entering the 
full NLO corrections to Higgs boson pair production. 
Therefore \pysecdec{} can open the door to the evaluation of higher order corrections to multi-scale processes which 
are not accessible by semi-analytical approaches.

\section*{Acknowledgements}
We would like to thank Viktor Papara, Rudi Rahn and Andries Waelkens for useful comments 
and Hjalte Frellesvig and Francesco Moriello for providing numbers for comparison.
This research was supported in part by the 
Research Executive Agency (REA) of the European Union under the Grant Agreement
PITN-GA2012316704 (HiggsTools) and the ERC Advanced Grant MC@NNLO (340983). 
S. Borowka gratefully acknowledges financial support by the ERC Starting Grant ``MathAm" (39568).

\renewcommand \thesection{\Alph{section}}
\renewcommand{\theequation}{\Alph{section}.\arabic{equation}}
\setcounter{section}{0}
\setcounter{equation}{0}


\section{Parameter settings}
\label{sec:appendix:defaults}

\subsection{Algebraic part}

The settings for the algebraic part are listed and explained in detail in the section 
``Code Writer/Make Package" in the documentation, 
as well as in the section ``Loop Integral" for settings which are specific to loop integrals.

\subsubsection{Loop package}

The parameters for {\tt loop\_package} are:
\begin{description}
\item[name]: string. The name of the {\it C++} namespace and the output directory.
    \item[loop\_integral]: The loop integral to be computed, defined via \\
    {\tt pySecDec.loop\_integral.LoopIntegral} (see below).
    \item[requested\_orders]: integer. The expansion in the regulator will be computed to this order.
    \item[real\_parameters]: iterable of strings or sympy symbols, optional. Parameters to be interpreted as real numbers, e.g. Mandelstam invariants and masses.
    \item[complex\_parameters]: iterable of strings or sympy symbols, optional. Parameters to be interpreted as complex numbers, e.g. masses in a complex mass scheme.
    \item[contour\_deformation (True)]: bool, optional. Whether or not to produce code for contour deformation.
    \item[additional\_prefactor (1)]: string or sympy expression, optional. An additional factor to be multiplied to the loop integral.
     It may depend on the regulator, the real parameters and the complex parameters.
    \item[form\_optimization\_level (2)]: integer out of the interval
      [0,3], optional. The optimization level to be used in \form{}. 
    \item[form\_work\_space ('500M')]: string, optional. The \form{} WorkSpace.
    \item[decomposition\_method]: string, optional. The strategy for decomposing the polynomials. The following strategies are available:
    \begin{itemize}
     \item   `iterative' (default)
      \item   `geometric'
      \item   `geometric\_ku'
   \end{itemize}
    \item[normaliz\_executable (`normaliz')]: string, optional. The command to run normaliz. normaliz is only required if {\tt decomposition\_method} is set to `geometric' or `geometric\_ku'. 
    \item[enforce\_complex (False)]: bool, optional. Whether or not the generated integrand functions should have a complex return type even though they might be purely real. The return type of the integrands is automatically complex if {\tt contour\_deformation} is True or if there are complex parameters. In other cases, the calculation can typically be kept purely real. Most commonly, this flag is needed if the 
    logarithm of a negative real number can occur in one of the integrand functions. 
    However, \pysecdec{} will suggest setting this flag to True in that case. 
    \item[split (False)]: bool, optional. Whether or not to split the integration domain in order to map singularities at 1 
    back to the origin. Set this option to True if you have singularities when one or more integration variables are 
    equal to one. 
  \item[ibp\_power\_goal (-1)]: integer, optional. The {\tt power\_goal} that is forwarded to the integration by parts routine. Using the default setting, integration by parts is applied until no linear or higher poles remain in the integral.  We refer to the documentation for more detailed information.
    \item[use\_dreadnaut (True)]: bool or string, optional. Whether or not to use {\tt dreadnaut} to find sector symmetries. 
\end{description}
The main keywords to define loop integrals from a ``graphical representation" ({\tt LoopIntegralFromGraph})
are:
\begin{description}
\item[internal\_lines]: list defining the propagators as connections between labelled vertices, 
where the first entry of each element denotes the mass of the propagator, e.g. {\tt [[`m', [1,2]], [`0', [2,1]]]}.
\item[external\_lines]: list of external line specifications, consisting of a string for the external momentum and a string or number labelling the vertex, e.g. {\tt [[`p1', 1], [`p2', 2]]}.
\item[replacement\_rules]: symbolic replacements to be made for the external momenta, e.g. definition of Mandelstam variables. Example: {\tt [(`p1*p2', `s'), (`p1**2', 0)]} where {\tt p1} and {\tt p2} are external momenta. It is also possible to specify vector replacements, e.g. {\tt [(`p4', `-(p1+p2+p3)')]}.
\item[Feynman\_parameters ('x')]: iterable or string, optional. The symbols to be used for the Feynman parameters. If a string is passed, the Feynman parameter variables will be consecutively numbered starting from zero.
\item[regulator ($\epsilon$)]: string or sympy symbol, optional. The symbol to be used for the dimensional regulator.
Note:
If you change this symbol, you have to adapt the dimensionality accordingly.
\item[regulator\_power (0)]: integer. The numerator will be multiplied by the regulator ($\eps$) raised to this power. 
This can be used to ensure that the numerator is finite in the limit $\eps\to 0$.
\item[dimensionality (4-2$\epsilon$)]: string or sympy expression, optional. The dimensionality of the loop momenta.
\item[powerlist]: iterable, optional. The powers of the propagators, possibly dependent on the regulator. The ordering must match the ordering of the propagators given in {\tt internal\_lines}.
\end{description}

For {\tt LoopIntegralFromPropagators}:
\begin{description}
\item[propagators]: iterable of strings or sympy expressions. The propagators in momentum representation, 
e.g. {\tt [`k1**2', `(k1-k2)**2 - m1**2']}.
\item[loop\_momenta]: iterable of strings or sympy expressions. The loop momenta, e.g. {\tt [`k1','k2']}.
\item[external\_momenta]:
    iterable of strings or sympy expressions, optional. The external momenta, e.g. {\tt [`p1','p2']}. 
    Specifying the external momenta is only required when a numerator is to be constructed.
\item[Lorentz\_indices]:
    iterable of strings or sympy expressions, optional. Symbols to be used as Lorentz indices in the numerator.
  \item[numerator (1)]: string or sympy expression, optional. The numerator of the loop integral. 
Scalar products must be passed in index notation, \\ 
e.g. {\tt k1(mu)*k2(mu)+p1(mu)*k2(mu)}. 
   All Lorentz indices  must be explicitly defined using the parameter {\tt Lorentz\_indices}.
 \item[metric\_tensor ('g')]: string or sympy symbol, optional. The symbol to be used for the (Minkowski) metric tensor $g^{\mu\nu}$.
\end{description}
Note: The parameters {\tt replacement\_rules, regulator, dimensionality, powerlist, regulator\_power} are available for both, 
{\tt LoopIntegralFromGraph} and {\tt LoopIntegralFromPropagators}.

\subsubsection{Make package}

The parameters for {\tt make package} are:
\begin{description}
\item[name]: string. The name of the {\it C++} namespace and the output directory.
    \item[integration\_variables]: iterable of strings or sympy symbols. The variables that are to be integrated from 0 to 1.
    \item[regulators]: iterable of strings or sympy symbols. The (UV/IR) regulators of the integral.
    \item[requested\_orders]: iterable of integers. Compute the expansion in the regulators to these orders.
    \item[polynomials\_to\_decompose]: iterable of strings or sympy expressions. The polynomials to be decomposed.
    \item[polynomial\_names]: iterable of strings. Assign symbols for the  polynomials to decompose. 
    These can be referenced in the {\tt other\_polynomials}.
    \item[other\_polynomials]:
    iterable of strings or sympy expressions. Additional polynomials where no decomposition is attempted. The symbols defined in {\tt polynomial\_names} can be used to reference the {\tt polynomials\_to\_decompose}. 
    This is particularly useful when computing loop integrals where the numerator can depend on the first and second Symanzik polynomials.
    Note that the {\tt polynomial\_names} refer to the {\tt polynomials\_to\_decompose} without their exponents.
    \item[prefactor]: string or sympy expression, optional. A factor that does not depend on the integration variables. It can depend on the regulator(s) and the kinematic invariants. The result returned by \pysecdec{} will contain the expanded prefactor.
    \item[remainder\_expression]:
    string or sympy expression, optional. An additional expression which will be considered as a multiplicative factor.
    \item[functions]: iterable of strings or sympy symbols, optional. Function symbols occurring in {\tt remainder\_expression}.
    Note:
    The power function pow and the logarithm log are already defined by default. The log uses the nonstandard continuation from a negative imaginary part on the negative real axis (e.g. $\log(-1) = -i\,\pi$).
  \item[form\_insertion\_depth (5)]: non-negative integer, optional. How deep FORM should try to resolve nested function calls. 
    \item[contour\_deformation\_polynomial]: string or sympy symbol, optional. The name of the polynomial in {\tt polynomial\_names} that is to be continued to the complex plane according to a $-i\delta$ prescription. 
    For loop integrals, this is the second Symanzik polynomial F, and this will be done automatically in {\tt loop\_package}. 
    If not provided, no code for contour deformation is created.
    \item[positive\_polynomials]: iterable of strings or sympy symbols, optional. The names of the polynomials in {\tt polynomial\_names} that should always have a positive real part. For loop integrals, this applies to the first Symanzik polynomial U. 
    If not provided, no polynomial is checked for positiveness. If {\tt contour\_deformation\_polynomial} is None, this parameter is ignored.
\end{description}
Note: All parameters (except {\tt loop\_integral}) described under {\tt loop\_package} are also available in {\tt make\_package}.

\subsection{C++ part}

The default settings for the numerical integration are listed in the section 
``Integral Interface" in the documentation. 
We also list the defaults and a short description for the main parameters here.
The values in brackets behind the keywords denote the defaults.

\subsubsection{Contour deformation parameters and general settings}

\begin{description}
\item[real\_parameters]: iterable of float. The real parameters of the library (e.g. kinematic invariants in the case of loop integrals).
\item[complex\_parameters]: iterable of complex. The complex parameters of the library (e.g. complex masses).
\item[together (True)]: bool. Determines whether to integrate the sum of all sectors or to integrate the sectors separately.
\item[number\_of\_presamples (100000)]: unsigned int, optional. The number of samples used for the contour optimization. This option is ignored if the integral library was created with contour deformation set to `False'. 
\item[deformation\_parameters\_maximum (1.0)]: float, optional. The maximal value the deformation parameters $\lambda_i$ can obtain. If number\_of\_presamples=0, all $\lambda_i$ are set to this value. This option is ignored if the integral library was created without deformation. 
\item[deformation\_parameters\_minimum ($10^{-5}$)]: float, optional. The minimal value for the deformation parameters $\lambda_i$. This option is ignored if the integral library was created without deformation. 
\item[deformation\_parameters\_decrease\_factor (0.9)]: float, optional. If the sign check (the imaginary part always must be negative) with the optimized $\lambda_i$ fails, all $\lambda_i$ are multiplied by this value until the sign check passes. This option is ignored if the integral library was created without deformation. 
\item[real\_complex\_together (False)]: If true, real and imaginary parts are evaluated simultaneously.  
If the grid should be optimally adapted to both real and imaginary part, it is more advisable to evaluate them separately.
\end{description}

\subsubsection{{\sc Cuba} parameters}

\begin{table}[h!]
\caption{Default settings for integrator-specific parameters.\label{tab:cuba}}
\begin{tabular}{|l|l|l|l|}
\hline
Vegas&Suave&Divonne&Cuhre\\
\hline
nstart (1000)&nnew (1000)&key1 (2000)&key (0)\\
nincrease (500)&nmin (10)&key2 (1), key3 (1)&\\
nbatch (1000)&flatness (25.0)&maxpass (4)&\\
&&border (0.0)&\\
&&maxchisq (1.0)&\\
&&mindeviation (0.15)&\\
\hline
\end{tabular}
\end{table}
 
Common to all integrators:
\begin{description}
\item[epsrel (0.01)]: The desired relative accuracy for the numerical evaluation. 
\item[epsabs ($10^{-7}$)]: The desired absolute accuracy for the numerical evaluation. 
\item[flags (0)]: Sets the {\sc Cuba} verbosity flags. 
The flags=2 means that the {\sc Cuba} 
input parameters and the result after each iteration are written to the log file of the 
numerical integration.
\item[seed (0)]: The seed used to generate random numbers for the numerical integration with Cuba.
\item[maxeval (1000000)]: The maximal number of evaluations to be performed by the numerical integrator.
\item[mineval (0)]: The number of evaluations which should at least be done before the numerical integrator returns a result. 
\end{description}
For the description of the more specific parameters, we refer to the
{\sc Cuba} manual. Our default settings are given in
Table~\ref{tab:cuba}.
When using Divonne, we strongly advise to use a non-zero value for
{\tt border}, e.g. $10^{-8}$.

 
\bibliographystyle{JHEP}

\bibliography{refs_secdec}

\providecommand{\href}[2]{#2}\begingroup\raggedright\begin{thebibliography}{10}

\bibitem{Henn:2013pwa}
J.~M. Henn, \emph{{Multiloop integrals in dimensional regularization made
  simple}}, \href{http://dx.doi.org/10.1103/PhysRevLett.110.251601}{\emph{Phys.
  Rev. Lett.} {\bfseries 110} (2013) 251601},
  [\href{https://arxiv.org/abs/1304.1806}{{\ttfamily 1304.1806}}].

\bibitem{Kotikov:1991pm}
A.~V. Kotikov, \emph{{Differential equation method: The Calculation of N point
  Feynman diagrams}}, \href{http://dx.doi.org/10.1016/0370-2693(91)90536-Y,
  10.1016/0370-2693(92)91582-T}{\emph{Phys. Lett.} {\bfseries B267} (1991)
  123--127}.

\bibitem{Gehrmann:1999as}
T.~Gehrmann and E.~Remiddi, \emph{{Differential equations for two loop four
  point functions}},
  \href{http://dx.doi.org/10.1016/S0550-3213(00)00223-6}{\emph{Nucl. Phys.}
  {\bfseries B580} (2000) 485--518},
  [\href{https://arxiv.org/abs/hep-ph/9912329}{{\ttfamily hep-ph/9912329}}].

\bibitem{Hepp:1966eg}
K.~Hepp, \emph{{Proof of the Bogolyubov-Parasiuk theorem on renormalization}},
  \href{http://dx.doi.org/10.1007/BF01773358}{\emph{Commun. Math. Phys.}
  {\bfseries 2} (1966) 301--326}.

\bibitem{Roth:1996pd}
M.~Roth and A.~Denner, \emph{{High-energy approximation of one loop Feynman
  integrals}},
  \href{http://dx.doi.org/10.1016/0550-3213(96)00435-X}{\emph{Nucl. Phys.}
  {\bfseries B479} (1996) 495--514},
  [\href{https://arxiv.org/abs/hep-ph/9605420}{{\ttfamily hep-ph/9605420}}].

\bibitem{Binoth:2000ps}
T.~Binoth and G.~Heinrich, \emph{{An automatized algorithm to compute infrared
  divergent multiloop integrals}},
  \href{http://dx.doi.org/10.1016/S0550-3213(00)00429-6}{\emph{Nucl. Phys.}
  {\bfseries B585} (2000) 741--759},
  [\href{https://arxiv.org/abs/hep-ph/0004013}{{\ttfamily hep-ph/0004013}}].

\bibitem{Heinrich:2008si}
G.~Heinrich, \emph{{Sector Decomposition}},
  \href{http://dx.doi.org/10.1142/S0217751X08040263}{\emph{Int. J. Mod. Phys.}
  {\bfseries A23} (2008) 1457--1486},
  [\href{https://arxiv.org/abs/0803.4177}{{\ttfamily 0803.4177}}].

\bibitem{Carter:2010hi}
J.~Carter and G.~Heinrich, \emph{{SecDec: A general program for sector
  decomposition}},
  \href{http://dx.doi.org/10.1016/j.cpc.2011.03.026}{\emph{Comput.Phys.Commun.}
  {\bfseries 182} (2011) 1566--1581},
  [\href{https://arxiv.org/abs/1011.5493}{{\ttfamily 1011.5493}}].

\bibitem{Borowka:2012yc}
S.~Borowka, J.~Carter and G.~Heinrich, \emph{{Numerical Evaluation of
  Multi-Loop Integrals for Arbitrary Kinematics with SecDec 2.0}},
  \href{http://dx.doi.org/10.1016/j.cpc.2012.09.020}{\emph{Comput.Phys.Commun.}
  {\bfseries 184} (2013) 396--408},
  [\href{https://arxiv.org/abs/1204.4152}{{\ttfamily 1204.4152}}].

\bibitem{Borowka:2015mxa}
S.~Borowka, G.~Heinrich, S.~P. Jones, M.~Kerner, J.~Schlenk and T.~Zirke,
  \emph{{SecDec-3.0: numerical evaluation of multi-scale integrals beyond one
  loop}}, \href{http://dx.doi.org/10.1016/j.cpc.2015.05.022}{\emph{Comput.
  Phys. Commun.} {\bfseries 196} (2015) 470--491},
  [\href{https://arxiv.org/abs/1502.06595}{{\ttfamily 1502.06595}}].

\bibitem{Soper:1999xk}
D.~E. Soper, \emph{{Techniques for QCD calculations by numerical integration}},
  \href{http://dx.doi.org/10.1103/PhysRevD.62.014009}{\emph{Phys. Rev.}
  {\bfseries D62} (2000) 014009},
  [\href{https://arxiv.org/abs/hep-ph/9910292}{{\ttfamily hep-ph/9910292}}].

\bibitem{Beerli:2008zz}
S.~Beerli, \emph{{A New method for evaluating two-loop Feynman integrals and
  its application to Higgs production}}.
\newblock PhD thesis, Zurich, ETH, 2008.

\bibitem{Bogner:2007cr}
C.~Bogner and S.~Weinzierl, \emph{{Resolution of singularities for multi-loop
  integrals}},
  \href{http://dx.doi.org/10.1016/j.cpc.2007.11.012}{\emph{Comput.Phys.Commun.}
  {\bfseries 178} (2008) 596--610},
  [\href{https://arxiv.org/abs/0709.4092}{{\ttfamily 0709.4092}}].

\bibitem{Gluza:2010rn}
J.~Gluza, K.~Kajda, T.~Riemann and V.~Yundin, \emph{{Numerical Evaluation of
  Tensor Feynman Integrals in Euclidean Kinematics}},
  \href{http://dx.doi.org/10.1140/epjc/s10052-010-1516-y}{\emph{Eur.Phys.J.}
  {\bfseries C71} (2011) 1516},
  [\href{https://arxiv.org/abs/1010.1667}{{\ttfamily 1010.1667}}].

\bibitem{Ueda:2009xx}
T.~Ueda and J.~Fujimoto, \emph{{New implementation of the sector decomposition
  on FORM}}, {\emph{PoS} {\bfseries ACAT08} (2008) 120},
  [\href{https://arxiv.org/abs/0902.2656}{{\ttfamily 0902.2656}}].

\bibitem{Kaneko:2010kj}
T.~Kaneko and T.~Ueda, \emph{{Sector Decomposition Via Computational
  Geometry}}, {\emph{PoS} {\bfseries ACAT2010} (2010) 082},
  [\href{https://arxiv.org/abs/1004.5490}{{\ttfamily 1004.5490}}].

\bibitem{Smirnov:2008py}
A.~Smirnov and M.~Tentyukov, \emph{{Feynman Integral Evaluation by a Sector
  decomposiTion Approach (FIESTA)}},
  \href{http://dx.doi.org/10.1016/j.cpc.2008.11.006}{\emph{Comput.Phys.Commun.}
  {\bfseries 180} (2009) 735--746},
  [\href{https://arxiv.org/abs/0807.4129}{{\ttfamily 0807.4129}}].

\bibitem{Smirnov:2009pb}
A.~Smirnov, V.~Smirnov and M.~Tentyukov, \emph{{FIESTA 2: Parallelizeable
  multiloop numerical calculations}},
  \href{http://dx.doi.org/10.1016/j.cpc.2010.11.025}{\emph{Comput.Phys.Commun.}
  {\bfseries 182} (2011) 790--803},
  [\href{https://arxiv.org/abs/0912.0158}{{\ttfamily 0912.0158}}].

\bibitem{Smirnov:2013eza}
A.~V. Smirnov, \emph{{FIESTA 3: cluster-parallelizable multiloop numerical
  calculations in physical regions}},
  \href{http://dx.doi.org/10.1016/j.cpc.2014.03.015}{\emph{Comput.Phys.Commun.}
  {\bfseries 185} (2014) 2090--2100},
  [\href{https://arxiv.org/abs/1312.3186}{{\ttfamily 1312.3186}}].

\bibitem{Smirnov:2015mct}
A.~V. Smirnov, \emph{{FIESTA4: Optimized Feynman integral calculations with GPU
  support}}, \href{http://dx.doi.org/10.1016/j.cpc.2016.03.013}{\emph{Comput.
  Phys. Commun.} {\bfseries 204} (2016) 189--199},
  [\href{https://arxiv.org/abs/1511.03614}{{\ttfamily 1511.03614}}].

\bibitem{Vermaseren:2000nd}
J.~A.~M. Vermaseren, \emph{{New features of FORM}},
  \href{https://arxiv.org/abs/math-ph/0010025}{{\ttfamily math-ph/0010025}}.

\bibitem{Kuipers:2013pba}
J.~Kuipers, T.~Ueda and J.~A.~M. Vermaseren, \emph{{Code Optimization in
  FORM}}, \href{http://dx.doi.org/10.1016/j.cpc.2014.08.008}{\emph{Comput.
  Phys. Commun.} {\bfseries 189} (2015) 1--19},
  [\href{https://arxiv.org/abs/1310.7007}{{\ttfamily 1310.7007}}].

\bibitem{Hahn:2004fe}
T.~Hahn, \emph{{CUBA: A library for multidimensional numerical integration}},
  \href{http://dx.doi.org/10.1016/j.cpc.2005.01.010}{\emph{Comput. Phys.
  Commun.} {\bfseries 168} (2005) 78--95},
  [\href{https://arxiv.org/abs/hep-ph/0404043}{{\ttfamily hep-ph/0404043}}].

\bibitem{Hahn:2014fua}
T.~Hahn, \emph{{Concurrent Cuba}},
  \href{http://dx.doi.org/10.1088/1742-6596/608/1/012066}{\emph{J. Phys. Conf.
  Ser.} {\bfseries 608} (2015) 012066},
  [\href{https://arxiv.org/abs/1408.6373}{{\ttfamily 1408.6373}}].

\bibitem{Schlenk:2016cwf}
J.~Schlenk and T.~Zirke, \emph{{Calculation of Multi-Loop Integrals with
  SecDec-3.0}},  in \emph{{Proceedings, 12th International Symposium on
  Radiative Corrections (Radcor 2015) and LoopFest XIV (Radiative Corrections
  for the LHC and Future Colliders): Los Angeles, CA, USA, June 15-19, 2015}},
  2016.
\newblock \href{https://arxiv.org/abs/1601.03982}{{\ttfamily 1601.03982}}.

\bibitem{Schlenk:2016a}
J.~Schlenk, \emph{{Techniques for higher order corrections and their
  application to LHC phenomenology}}.
\newblock PhD thesis, Technical University Munich, 2016.

\bibitem{Kaneko:2009qx}
T.~Kaneko and T.~Ueda, \emph{{A Geometric method of sector decomposition}},
  \href{http://dx.doi.org/10.1016/j.cpc.2010.04.001}{\emph{Comput.Phys.Commun.}
  {\bfseries 181} (2010) 1352--1361},
  [\href{https://arxiv.org/abs/0908.2897}{{\ttfamily 0908.2897}}].

\bibitem{2013arXiv1301.1493M}
B.~D. {McKay} and A.~{Piperno}, \emph{{Practical graph isomorphism, II}},
  {\emph{ArXiv e-prints} (Jan., 2013) },
  [\href{https://arxiv.org/abs/1301.1493}{{\ttfamily 1301.1493}}].

\bibitem{graphviz}
http://www.graphviz.org.

\bibitem{2012arXiv1206.1916B}
W.~{Bruns}, B.~{Ichim} and C.~{S{\"o}ger}, \emph{{The power of pyramid
  decomposition in Normaliz}}, {\emph{ArXiv e-prints} (June, 2012) },
  [\href{https://arxiv.org/abs/1206.1916}{{\ttfamily 1206.1916}}].

\bibitem{Normaliz}
W.~{Bruns}, B.~{Ichim}, T.~{R{\"o}mer} and C.~{S{\"o}ger}, ``{{Normaliz.
  Algorithms for rational cones and affine monoids. Available from
  http://www.math.uos.de/normaliz}}.''

\bibitem{Fleischer:1997bw}
J.~Fleischer, A.~V. Kotikov and O.~L. Veretin, \emph{{The Differential equation
  method: Calculation of vertex type diagrams with one nonzero mass}},
  \href{http://dx.doi.org/10.1016/S0370-2693(97)01195-7}{\emph{Phys. Lett.}
  {\bfseries B417} (1998) 163--172},
  [\href{https://arxiv.org/abs/hep-ph/9707492}{{\ttfamily hep-ph/9707492}}].

\bibitem{Davydychev:2003mv}
A.~I. Davydychev and M.~Kalmykov, \emph{{Massive Feynman diagrams and inverse
  binomial sums}},
  \href{http://dx.doi.org/10.1016/j.nuclphysb.2004.08.020}{\emph{Nucl.Phys.}
  {\bfseries B699} (2004) 3--64},
  [\href{https://arxiv.org/abs/hep-th/0303162}{{\ttfamily hep-th/0303162}}].

\bibitem{Bonciani:2003hc}
R.~Bonciani, P.~Mastrolia and E.~Remiddi, \emph{{Master integrals for the two
  loop QCD virtual corrections to the forward backward asymmetry}},
  \href{http://dx.doi.org/10.1016/j.nuclphysb.2004.04.011}{\emph{Nucl.Phys.}
  {\bfseries B690} (2004) 138--176},
  [\href{https://arxiv.org/abs/hep-ph/0311145}{{\ttfamily hep-ph/0311145}}].

\bibitem{Ferroglia:2003yj}
A.~Ferroglia, M.~Passera, G.~Passarino and S.~Uccirati, \emph{{Two loop
  vertices in quantum field theory: Infrared convergent scalar
  configurations}},
  \href{http://dx.doi.org/10.1016/j.nuclphysb.2003.12.016}{\emph{Nucl.Phys.}
  {\bfseries B680} (2004) 199--270},
  [\href{https://arxiv.org/abs/hep-ph/0311186}{{\ttfamily hep-ph/0311186}}].

\bibitem{Heinrich:2007at}
G.~Heinrich, T.~Huber and D.~Maitre, \emph{{Master integrals for fermionic
  contributions to massless three-loop form-factors}},
  \href{http://dx.doi.org/10.1016/j.physletb.2008.03.028}{\emph{Phys. Lett.}
  {\bfseries B662} (2008) 344--352},
  [\href{https://arxiv.org/abs/0711.3590}{{\ttfamily 0711.3590}}].

\bibitem{vonManteuffel:2015gxa}
A.~von Manteuffel, E.~Panzer and R.~M. Schabinger, \emph{{On the Computation of
  Form Factors in Massless QCD with Finite Master Integrals}},
  \href{http://dx.doi.org/10.1103/PhysRevD.93.125014}{\emph{Phys. Rev.}
  {\bfseries D93} (2016) 125014},
  [\href{https://arxiv.org/abs/1510.06758}{{\ttfamily 1510.06758}}].

\bibitem{Bonciani:2016qxi}
R.~Bonciani, V.~Del~Duca, H.~Frellesvig, J.~M. Henn, F.~Moriello and V.~A.
  Smirnov, \emph{{Two-loop planar master integrals for Higgs$\to 3$ partons
  with full heavy-quark mass dependence}},
  \href{http://dx.doi.org/10.1007/JHEP12(2016)096}{\emph{JHEP} {\bfseries 12}
  (2016) 096}, [\href{https://arxiv.org/abs/1609.06685}{{\ttfamily
  1609.06685}}].

\bibitem{Primo:2016ebd}
A.~Primo and L.~Tancredi, \emph{{On the maximal cut of Feynman integrals and
  the solution of their differential equations}},
  \href{http://dx.doi.org/10.1016/j.nuclphysb.2016.12.021}{\emph{Nucl. Phys.}
  {\bfseries B916} (2017) 94--116},
  [\href{https://arxiv.org/abs/1610.08397}{{\ttfamily 1610.08397}}].

\bibitem{Fleischer:1998nb}
J.~Fleischer, A.~V. Kotikov and O.~L. Veretin, \emph{{Analytic two loop results
  for selfenergy type and vertex type diagrams with one nonzero mass}},
  \href{http://dx.doi.org/10.1016/S0550-3213(99)00078-4}{\emph{Nucl. Phys.}
  {\bfseries B547} (1999) 343--374},
  [\href{https://arxiv.org/abs/hep-ph/9808242}{{\ttfamily hep-ph/9808242}}].

\bibitem{Dubovyk:2016zok}
I.~Dubovyk, A.~Freitas, J.~Gluza, T.~Riemann and J.~Usovitsch, \emph{{30 years,
  some 700 integrals, and 1 dessert, or: Electroweak two-loop corrections to
  the Z$\bar b$b vertex}}, {\emph{PoS} {\bfseries LL2016} (2016) 075},
  [\href{https://arxiv.org/abs/1610.07059}{{\ttfamily 1610.07059}}].

\bibitem{Huber:2005yg}
T.~Huber and D.~Maitre, \emph{{HypExp: A Mathematica package for expanding
  hypergeometric functions around integer-valued parameters}},
  \href{http://dx.doi.org/10.1016/j.cpc.2006.01.007}{\emph{Comput. Phys.
  Commun.} {\bfseries 175} (2006) 122--144},
  [\href{https://arxiv.org/abs/hep-ph/0507094}{{\ttfamily hep-ph/0507094}}].

\bibitem{Binoth:2002xg}
T.~Binoth, E.~W.~N. Glover, P.~Marquard and J.~J. van~der Bij, \emph{{Two loop
  corrections to light by light scattering in supersymmetric QED}},
  \href{http://dx.doi.org/10.1088/1126-6708/2002/05/060}{\emph{JHEP} {\bfseries
  05} (2002) 060}, [\href{https://arxiv.org/abs/hep-ph/0202266}{{\ttfamily
  hep-ph/0202266}}].

\end{thebibliography}\endgroup

\end{document}